# From fifty years ago, the birth of modern liquid-state science

## David Chandler[1]

[1]Department of Chemistry, University of California, Berkeley, CA, 94720; email: chandler@berkeley.edu

**Abstract**

The story told in this autobiographical perspective begins fifty years ago at the 1967 Gordon Research Conference on the Physics and Chemistry of Liquids. It traces developments in liquid-state science from that time, including contributions from the author, and especially in the study of liquid water. It emphasizes the importance of fluctuations, and the challenges of far-from-equilibrium phenomena.

## Contents

### 1. Introduction

In the summer of 1967, I was 22 years old, at the end of my first year at Harvard Graduate School. I had just joined Roy Gordon's research group, and Roy brought me along to the Gordon Research Conference (GRC) on Liquids. I was one of only three graduate students attending. GRCs are prestigious scientific meetings, conceived in the 1930s by chemist Neil Gordon (no relationship to Roy). Participation was regarded as an activity for specialists only, and I felt it was a special honor to be able to meet and talk with all the leading experts. You can find me in **Figure 1**, pictured at the left end of the second row. The other two graduate students were Harry Swinney (three to my left in the picture), who was studying with Herman Cummins at CUNY, and Charles Bennett (near the middle of the picture in a plaid shirt), who was studying with David Turnbell at Harvard. Harry was to become famous for his systematic experiments on liquid turbulence and chaos. Charles also was to become famous, first for his contributions to classical molecular simulations (e.g., rare event and free energy computations), and then as a founder of quantum information theory.

> Understanding liquid matter at a molecular level was to be a non-trivial and significant undertaking, and in 1967 the time was ripe for it.

The topic of the meeting, the physics of liquids, was a challenging and central subject for those interested in statistical mechanics. The great Russian physicist, Lev Landau, had written "Unlike solids and gases, liquids do not allow a general calculation of their thermodynamic quantities or even their temperature dependence."(1) This pessimism was rooted in the fact that liquids are strongly interacting but disordered systems whose very existence implies a delicate balance between energy and entropy. Understanding liquid matter at a molecular level was to be a non-trivial and significant undertaking. The time was ripe for that undertaking. Experiments such as neutron scattering, laser light scattering and pressure tuning spectroscopy were being done for the first time. Nascent theoretical

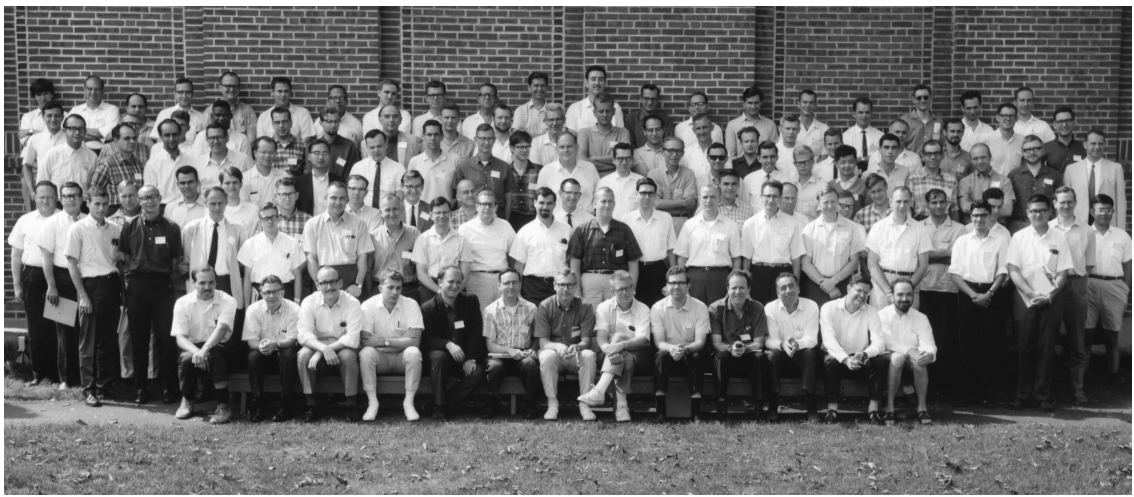

**Figure 1**

Attendees at the Gordon Research Conference on Physics and Chemistry of Liquids, Summer 1967, held at Procter Academy, Andover, New Hampshire, and chaired by Marshall Fixman. (Photograph provided courtesy of Gordon Research Conferences.)

interpretations were being developed using correlation functions, and the first molecular simulations were becoming available to test the underpinnings of these theories.

Marshall Fixman, seated in the middle of the front row, chaired the meeting. Robert Zwanzig and then Roy Gordon are immediately to his left. Ben Widom and then Leo Kadanoff are immediately to his right. True to Fixman's greatest interests at the time, the meeting featured much discussion of critical phenomena. It also featured discussion about the structure and dynamics of simple liquids. The foundations for understanding both topics would soon be in hand — two scientific revolutions, one spectacular the other quiet but still profound. In these pages, I describe these developments and the people who made them happen, almost all of whom I came to meet for first time at that GRC 50 years ago.

## 2. Gordon Research Conferences

A notable feature in **Figure** 1 is the nearly ubiquitous pen holders in men's breast pockets. Fountain pens often leaked! Another is the business-like attire, in contrast to the standard shorts and t-shirts of today's GRCs. And most striking, there is not a single woman participant. A few years later, women began to populate the meeting. For example, I first met Sandra Greer, Branka Ladanyi and J. M. H. ('Anneke') Levelt Sengers at the 1973 GRC, scientists I have worked with and enjoyed knowing for many years after. All three have had distinguished careers, Sandra and Anneke for pioneering experiments on critical phenomena, and Branka for influential computations on solvation dynamics.

The general philosophy of a GRC goes back to its founding meetings in the 1930s: sequestered venues where experts have plenty of opportunity to present and discuss new and even tentative ideas; and only minimal discussion of published work, because experts should already be familiar with that material. The morning and evening lecture sessions are to be intense, while afternoon recreation, facilitated by the typical rural settings at private New England boarding schools, allows for relaxed encounters between participants.

In 1969, the Liquids GRC moved north from Proctor Academy, to settle at the Holderness School in Plymouth, New Hampshire, located in the foothills of the White Mountains about seven miles from Squam Lake. I find the area enchanting, in no small part because the territory reminds me of my childhood summers spent 40 miles to the north-east, on Long Lake, in Maine, where I learned camping, swimming, sailing and tennis. Starting in 1975, my family and I have spent a month or more nearly every summer in New Hampshire, eventually purchasing a family vacation property on Squam Lake, near the original location of the Holderness School.

Since 1969, the Holderness School has nearly doubled in size. The character of the Liquids GRC has changed too. Most significantly, many graduate students now typically populate the meeting, review lectures are not uncommon, and question periods and dinner-table conversation seem less focused than I remember from 1967. Bringing new students into the fold is important, but too large a meeting might diminish what was the unique role of GRCs.

## 3. Criticality

Along with Marshall Fixman, four other intellectual giants of critical phenomena were in attendance — Michael Fisher and Lars Onsager, in addition to Kadanoff and Widom. As noted, Kadanoff and Widom are seated in the front row to Fixman's right, but Fisher and Onsager are not featured among the pictured participants. They awakened too late to be included.

I had first set eyes on Fisher and Onsager a few months earlier, at the spring gathering of Joel Lebowitz's semi-annual statistical mechanics meeting. Joel is pictured seated at the right end of the front row. His meetings continue on as major events twice every year at Rutgers University in New Jersey, but in the 1960s, the venue for the meeting was a classroom at Yeshiva University in New York City. At the time, Lebowitz was on the faculty of Yeshiva's Belfer Graduate School of Science. (With this meeting in mind, I named Berkeley's annual meeting the *Mini* Stat Mech Meeting, to distinguish it from Joel's longer-running stat mech meeting.) In those early years, everyone attending the Yeshiva meeting was expected to present a four- to five-minute talk on some new result or idea, using a blackboard for no more than one equation or sketched graph. I recall that Michael Fisher had something intelligent (and often critical) to say about nearly every presentation. He was intimidating.

Onsager, on the other hand, was a quiet presence. In fact, he was napping in the chair next to mine for most of the morning at that Yeshiva meeting, and I was surprised to learn, when his name was called, that this old man was Onsager himself. (It was a year prior to him being awarded the Nobel Prize for his contributions to non-equilibrium statistical mechanics.) Onsager went to the board, mumbled something as he wrote an equation blocked from view by his notable height and broad shoulders. Upon completing the equation, he turned to show us all, and with a smile and a twinkle in his eye, he told us the uncovered formula was a remarkable result. I couldn't understand its content, but the presentation so captivated me that to this day I almost always think of Onsager when

I say or write "remarkable."

In current times, we all recognize their names for what these men accomplished on critical phenomena. The modern field was initiated by Onsager's exact solution of the two-dimensional Ising model (2). The concept of universality and scaling began with Widom's homogeneity hypothesis for thermodynamics close to a critical point (3), followed by Kadanoff's derivation and generalization of Widom's hypothesis, accounting for spatial scaling of correlated fluctuations close to a critical point (4). The culmination of these ideas came in 1971 with Kenneth Wilson's Nobel-Prize-winning renormalization group theory, a discovery in which Michael Fisher played a hugely important supporting role (5, 6, 7). Along the way, Fixman (8) generalized the Ornstein-Zernike model of density fluctuations, and Fisher (9) introduced the exponent $\eta$, setting the stage for the important concept of fractional dimensionality.

During the 1965-66 Academic Year, as an MIT student doing an undergraduate research thesis with Irwin Oppenheim, I studied the scaling papers of Widom and Kadanoff, the 1964 review by Fisher, the Ising model paper by Onsager, and an assortment of papers by Fixman, including three with his student W. Botch. (I would chuckle thinking of Fixman fixing botched calculations.) It was the year I fell in love with scientific research, and especially with theory. I had much to occupy my attention, with advanced courses in quantum mechanics, classical mechanics and statistical mechanics, plus two humanities classes and college tennis. In the midst of all that, I set aside two nights every week for studying the topic assigned to me by Professor Oppenheim — the statistics of density fluctuations near a critical point. I always looked forward to those evenings. I learned from the literature, and I tried to be inventive. Trying to be inventive gave me a thrill. Playing with equations helped me develop a deep understanding of basic concepts. It gave me a sense of the theoretical structure surrounding those concepts, and encouragement from Professor Oppenheim built my confidence.

I am sometimes asked how I developed an ability to be original in scientific research. It is those evenings at MIT, playing with equations, when I acquired the taste for discovery. Importantly, I have always avoided projects that involve calculations I already understand. The avoidance made me a poor research assistant during my graduate and postdoctoral years. (I never published a paper with either my Ph.D. advisor or my postdoctoral advisor.) My interest is sparked when a subject confuses me. Usually the confusion is the result of my ignorance, but sometimes, after enough thought and investigation, I learn that the essence of the confusion is something fundamental. Resolving that confusion becomes a quest, and good science often results.

> Elucidating consequences of spatial self-similarity, Kadanoff formulated the physical perspective from which renormalization group theory emerged.

Having studied aspects of critical phenomena, I was primed to hear the GRC lectures by Kadanoff and Widom. Widom presented his analysis of how the direct correlation function becomes long-ranged at the critical point (a consequence of $\eta$), and Kadanoff presented his perspective of renormalizing and self-similarity through block averaging of spins (see **Figure 2**). Both were brilliant and inspiring. Kadanoff's ideas, in particular, were setting the stage for the renormalization group theory. All the while, Michael Fisher sat in the audience

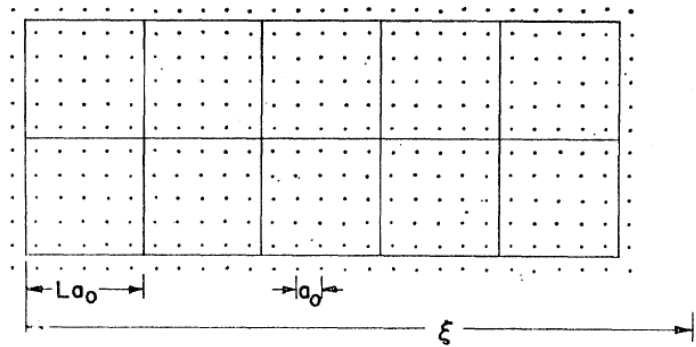

**Figure 2**

Kadanoff's division of Ising model into blocks, with $L \gg 1$, but $La_0 \ll \xi$ where $\xi$ is the correlation length of the model. From Kadanoff *et al.* (10). An Ising model with a large correlation length is an Ising model close to criticality.

nodding, which caused Kadanoff to exclaim, "Michael seems to approve, so I remain on safe ground." Onsager seemed to nap but somehow had interesting things to say at the end of each lecture. (Leo Kadanoff passed away in 2015, but a YouTube record of him speaking, https://www.youtube.com/watch?v=8an6x-NLfIc, captures some of his style, a talent for exposition that will be missed.)

From those studies of criticality emerged fundamental concepts of contemporary liquid-state science — statistical field theory, order parameters, broken symmetries, scaling, wetting and so forth. These are some of the important tools with which we now describe the behaviors of complex inhomogeneous fluids, systems for which liquid interfaces and large fluctuations are ubiquitous. Space-time versions of these concepts and techniques are also central in describing systems far from equilibrium such as glass formers and active matter. The topics discussed at contemporary Liquids GRCs are now far broader than in the 1960s and 70s largely because the tools created to study criticality opened the way for studying complex systems in general.

### 4. Molecular simulation

Computer experiments and simulations are another set of essential tools for today's science. But in 1967, few researchers had the resources to carry out such numerical work, and few had the imagination to recognize its advantages. Scientists at Los Alamos, including Enrico Fermi, Nick Metropolis and Edward Teller, were the first, e.g., (11) and (12). Four pioneers of the approach attended the 1967 GRC — Bill Wood (pictured in the second row, four to the right of me), Berni Alder (seated third from right in the front row), Loup Verlet and Aneesur Rahman (standing together, third and fourth from the right in the third row).

Wood and Alder made their first important contributions with machine calculations a few years earlier, establishing that a sufficiently compressed fluid of hard spheres will crystalize by a first-order phase transition (13, 14). Wood worked at Los Alamos Laboratory and used the Monte Carlo technique. Alder worked at Livermore Laboratory and applied

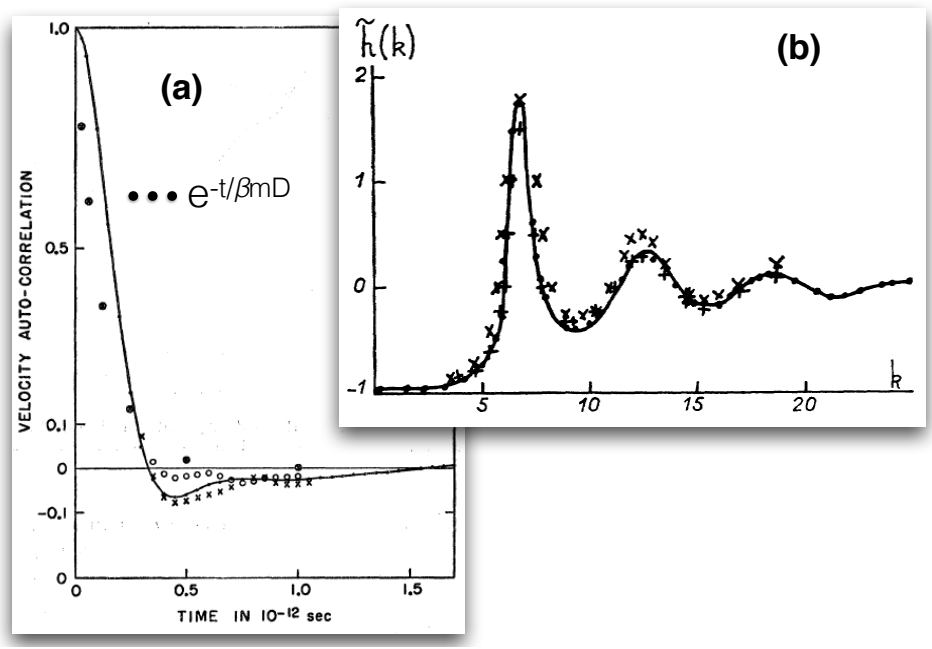

**Figure 3**

Results from early computer simulations of liquid argon near its triple point. (a) The velocity autocorrelation function computed from molecular dynamics of the Lennard-Jones model (solid line, for which open circles and crosses indicate error estimates). Its non-monotonic time dependence contrasts with the monotonic exponential function of Langevin theory with same diffusion constant (black dots). Here, $\beta = 1/k_\text{B}T$, $D$ is the self-diffusion constant and $m$ is particle mass. From Rahman (16). (b) The structure factor of liquid argon computed from molecular dynamics of the Lennard-Jones model (black dots), compared with that of the hard-sphere model with an empirically chosen hard-sphere diameter (solid line), and those measured with neutron and x-ray scattering (+'s and ×'s). The wavevector, $k$, is in units of the inverse Lennard-Jones particle diameter. From Verlet (17).

the molecular dynamics method. Alder's group initially encountered difficulties establishing the transition, and for a while there was some speculation from notable figures like John Kirkwood that Monte Carlo algorithms might be flawed. Eventually, the problem was tracked down to poor equilibration in the molecular dynamics calculation, and consistency was found when both methods were initiated from the same configuration taken from an equilibrated system (15).

At high enough packing fraction, the crystal state becomes the stable state signifying that the entropy of the crystal is *higher* than that of the amorphous fluid or disordered solid. (Recall that there is no average potential energy in the hard-sphere system, just entropy, so the relevant free energy is entirely the temperature-times-entropy contribution, $-TS$.) This seemingly counter intuitive result reflects the fact that each particle has free volume to vibrate in the ordered array while a significant fraction of particles are jammed in the disordered array. It is not simply a theoretical curiosity because freezing of real materials is often due to the same mechanism. But establishing that fact required the big step from computation to simulation. This step was initiated by Anees Rahman (16) and soon after elaborated by Loup Verlet (17).

Before Verlet's work, Ben Widom (18) had already provided physical arguments suggesting that the freezing of real materials was essentially the freezing of hard spheres. I will comment on these arguments, but first focus on the difference between computation and simulation. Before Rahman, it was understood that the statistical properties of a many-particle system could be learned from numerical experiments. But the systems studied in that way — hard spheres, chains of oscillators, and so forth — were viewed as instructive models, not as accurate approximations to natural materials. Alder even went as far to assert, incorrectly, that it would be impractical to attempt a reasonable level of realism in a machine calculation (19).

Rahman's insight was to approach nature on its own terms, and to build an algorithm to mimic natural trajectories (16). One striking and influential result obtained in this way is the degree to which liquid argon's velocity autocorrelation function exhibits non-monotonic behavior. See **Figure 3a**. (Hard spheres have much smaller negative-velocity correlations. The more interesting result for hard spheres is the long-time power-law decay of positive velocity correlations (20) manifesting a coupling between collective hydrodynamic modes and the velocity of the tagged particle.) With Rahman's simulations, we had for the first time precise "experimental" results for correlation functions of a realistic liquid where the microscopic interactions and laws of motion were completely determined. This lack of ambiguity is the ultimate challenge to theorists — no arbitrary fitting parameters allowed! Asked at the GRC about the cost of his calculations, Rahman demurred by saying it was trivial in comparison to one hour's expense in the ongoing Vietnam War. Nowadays, after decades of exponential decrease in cost per computing cycle (Moore's law), computing cost is not generally a relevant variable.

> Computer simulation became a new form of experimentation, with omnipotent and omniscient experimenters.

Verlet seized upon Rahman's work, significantly improving its algorithmic efficiencies by introducing the Verlet integrator and the Verlet neighbor list (17). He was thereby able to significantly extend what could then be accomplished with the approach. Results he presented at the 1967 GRC (e.g., **Figure 3b**) concluded with the demonstration that "most of [liquid argon] structure ... at high density is due to the geometrical effects produced by the existence of strong [interparticle repulsions] and that freezing transition of the model hard-sphere gas is related to the actual liquid-solid transition."

## 5. WCA Theory

Ben Widom connected the freezing of hard spheres to that of simple liquids like argon and xenon through a principle of corresponding states (18). He justified the principle with a compelling physical argument for how attractive forces tend to cancel due to packing constraints at the high densities where these liquids freeze. In 1969-70, when John Weeks and I worked together as postdoctoral researchers in Kurt Shuler's group at UC San Diego,

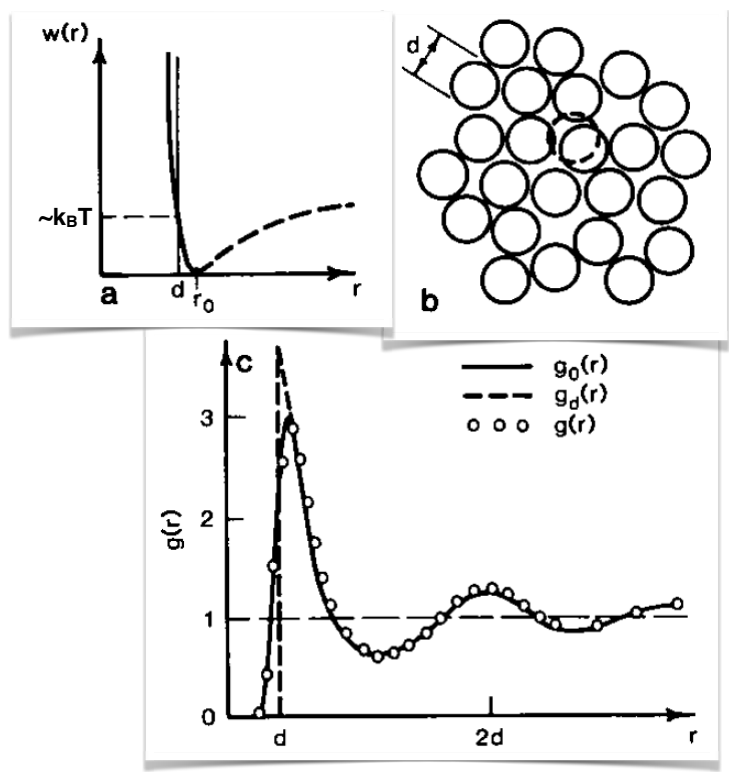

**Figure 4**

Repulsive forces and packing dominate the structure of a simple liquid. (a) Pair potential $w(r)$, its repulsive branch $w_0(r)$ (the solid line), its attractive branch (the dashed line), and the associated hard-sphere potential, $w_d(r)$ (the vertical line). (b) Schematic view of a region of a liquid composed of spherical particles interacting via a pair potential $w(r)$. (c) Radial distribution function, $g(r)$, of the liquid near its triple point (circles), of the reference fluid with only the repulsive forces (solid line), and of the associated hard-sphere fluid (dashed line). The results are captured by the principal equations from WCA theory:

$$g(r) \approx g_0(r) \approx g_d(r) \exp\{-[w_0(r) - w_d(r)]/k_{\rm B}T\}.$$

From Chandler *et al.* (22).

we often talked about Widom's argument and Verlet's results, especially during times when Hans Andersen was visiting. (Hans was then an Assistant Professor at Stanford, and he would visit San Diego on occasion to write papers with me on research he and I did during my graduate work at Harvard, when Hans was a Harvard Junior Fellow.) The Weeks-Chandler-Andersen theory (21) grew out of those discussions.

**Figure 4** illustrates the basic idea of that theory, the essential feature being that attractive forces act over much larger length scales and are much more slowly varying than repulsions. As a result, at the high densities of a typical liquid at standard conditions, repulsions will dominate the response to typical particle displacements (like that illustrated with the dashed circle). The attractive forces acting on a given particle will come from many surrounding particles and will thus tend to cancel, leaving only a constant or mean potential energy field. That mean field makes the high density liquid stable at low pressures while the inter-particle repulsions — packing forces — control fluctuations. In effect, therefore, Landau's concern over complexity due to competition between entropy and energy is resolved because one of the two players, the energy, can be treated by neglecting its fluctuations. Moreover, the fluctuations that do occur are only small in amplitude, which makes them easy to treat analytically.

> Repulsive forces, which are almost hard-sphere like, are found to determine the structure of dense liquids.

The equations that capture these ideas are remarkably simple. A temperature-dependent hard-sphere reference system emerges, and a first-order perturbation expression for thermodynamics is exceedingly accurate. Prior to this work, perturbation theories of liquids had been proposed by many — Robert Zwanzig, Joel Lebowitz, Stuart Rice, John Barker and Douglas Henderson, to name a few. But none of these earlier theories resulted in the simplicity, utility and physical picture that arose so naturally from following the path of applying Widom's compelling argument and explaining Verlet's results.

As sensible as that theory now seems, it took over a year of wrangling with referees before the principal WCA paper was accepted by the *Journal of Chemical Physics* (21). Fortunately, Editor Willard Stout sensed the significance of the work and continued discussion until a reasonable outcome could result. Soon after, the community agreed with our approach, and Hans was invited to speak about it at the 1973 Liquids GRC. In recent years, the WCA perspective continues to be relevant because simple liquids have returned to center stage in research on structural glass formers. Packing forces are largely responsible for the correlated dynamics and local rigidity characterizing such materials. Curiously, however, the lessons of that theory are sometimes ignored. Analysis and underlying physical principles are no longer thoroughly taught to researchers using computer simulations.

## 6. Time-correlation functions

A few years prior to Rahman's 1964 paper, time-correlation functions became a widely used tool. A decade earlier, Herbert Callen and Theodore Welton derived the fluctuation-dissipation theorem (23), and Mel Green (24) and Ryogo Kubo (25) derived formulas for transport coefficients. (Mel Green is pictured in the middle of the fourth row of 1967 GRC participants, flanked by Charles Bennett to his right, and Bob Pecora to his left.) In the early 1930s, Onsager had expressed the content of the fluctuation-dissipation theorem in his Nobel Prize winning papers (26, 27), but Onsager's cryptic style kept most scientists uninformed until Callen and Welton's work. (At the 1968 Nobel Prize Awards Ceremony, the Royal Swedish Academy of Sciences remarked on Onsager's brevity and his being "ahead of his time.") Onsager's remarkable insight relates non-equilibrium phenomena to time correlations of spontaneous fluctuations at equilibrium. It remained the only general and systematic formalism for more than a half-century. Only in the last decade, with the development of far-from-equilibrium fluctuation theorems (28, 29) and large-deviation theory (30), has a more generally applicable methodology appeared.

Bruce Berne, pictured in the second row directly behind Fixman and Widom, was one of the first students of what was then the new theories of time correlation functions. Berne had done his Ph.D. work under Stuart Rice's guidance at the University of Chicago. With Jean-Pierre Boon and Rice (31), he provided the first successful theoretical interpretation of Rahman's nonmonotonic velocity autocorrelation function. Indeed, Rahman's result stimulated the development of Berne's approach, which is a memory-function theory. Unlike the Langevin theory, which pictures a particle buffeted by random uncorrelated forces with a white noise spectrum, a memory-function adds color, implying random forces at one point in time are correlated to (i.e., remember) those at earlier times. In effect, a tagged particle recoils from its movement towards a group of neighboring particles because it takes a finite time for these neighbors to reorganize in response to the tagged particle. This finite time is the random force memory time, and the recoil produces non-monotonic decay of velocity correlations.

> Memory of fluctuations at equilibrium determines decay of time correlations and irreversibility close to equilibrium.

Projection operator formalisms of Zwanzig (32) and Hazime Mori (33) offer the underlying theory from which this picture is derived, and also by which mode coupling theories are derived. More generally, the formalisms played important roles in clarifying the physics of irreversibility. Prior to their derivations in the early 1960s, a major question was how irreversibility could emerge from a microscopically reversible dynamics. There were vague statistical arguments going back at least to the time of Ludwig Boltzmann, but an explicit and useful quantitative explanation was missing. Ilya Prigogine offered explanations based upon complicated but approximately truncated expansions of the classical Liouville's equation (34). The approach proved less than compelling if not incomprehensible.

What changed with Zwanzig and Mori was the idea of partitioning many body dynamics into a primary part (that is directly observable) and a secondary part (like that of a bath). Energy flows between the two parts, providing the mechanism of dissipation. Projection operators provide a convenient and exact bookkeeping device for carrying out the partitioning. The memory function emerges as the time correlation function for the fluctuations of forces of the bath coupled to the primary degrees of freedom. To the extent that the bath relaxes quickly compared to the primary variables, the formalism reduces to the Langevin theory with instantaneous memory loss (i.e., white noise) and exponential decay of velocity correlations.

During a breakfast table conversation at the 1967 GRC, I overheard Bruce Berne talking to Leo Kadanoff about these ideas and his theory of the velocity autocorrelation function. But at the time, the conversation was over my head. My first understanding came that fall, when I took a statistical mechanics course from Paul Martin in the Harvard Physics Department. Martin covered material he was assembling for a short monograph (35), which included topics he and Kadanoff had investigated on the time-correlation function analysis of hydrodynamics (36). (Kadanoff had been Martin's student. Kadanoff (37) attributed his work with Martin as providing the perspective that guided his initial contributions to critical phenomena.) With Martin's course under my belt, I was prepared to understand the papers by Mori, by Zwanzig and by Berne. A few years later, Bruce Berne published a pedagogical development of Mori's and Zwanzig's ideas (38), and I always enjoy referring to this development when I teach this material.

Like WCA theory of liquid structure, Berne's theory of the velocity autocorrelation function succeeded at finding a theoretical model that succinctly and correctly summarizes the underlying physics of a significant computer simulation result. It is a style that is sometimes overlooked in modern theoretical chemistry, which often relies too heavily on numerical work. Only a theoretical model, tested against numerical simulation, leads to understanding how complex systems work. The theoretical model can be (and should be) very simple because molecular simulation removes the need for complicated theory.

## 7. Water

When John Weeks and I embarked on WCA calculations in 1970, our supervisor Kurt Shuler pointed out that our postdoctoral salaries were paid by a grant from the US Office of Navel Research, and "they don't float boats on liquid argon." This is not to say that Kurt wasn't supportive and ultimately proud that members of his group formulated WCA theory. But at its start, he was understandably concerned that two youngsters might be venturing off on a foolish quest to do better on a topic where more senior and established researchers had already done much. And treating water, the liquid on which boats do float, was beyond anyone's capacity.

> Water, the liquid of life, has intrinsic complexity not present in simple liquids like argon.

Water is the welcoming host for molecules of many types. Its phase behavior, its electrostatics and its chemistry are sources of molecular driving forces for all life on Earth. It is much more complicated than argon, in part because the energetics of hydrogen bonding is equal in importance to entropy in determining its intermolecular correlations. At the time, in 1970, there were solid ideas about water's molecular nature from the works of J. D. ('Sage') Bernal (39) and from John Pople (40) who argued that "water is a broken-down form of the ice lattice, in which [hydrogen-bond] association extends throughout the whole liquid." But these were rare examples in a field dominated by speculation, much of it bizarre by today's standards. A commonly invoked two-state model required picturing liquid water as a mixture of gas and icebergs! (Once, in 1972, when I questioned one of these implied pictures at an early GRC on Water and Aqueous Solutions, I was called a "whipper snapper" and told to be quiet.)

The state of understanding was improved by Frank Stillinger's systematic theoretical studies of water (41). He ingeniously built reasonable empirical force field models (the ST2 pair potential is one example), and he enticed Anees Rahman to carry out simulations with the models, the first announced in a 1971 publication (42). (At the 1967 GRC, Frank is pictured in the second row, behind Peter Egelstaff who sits between Berni Alder and Joel Lebowitz.) Rahman's calculations were not the first attempts at simulating liquid water, but thanks to Stillinger's modeling, they were the first to capture a plausible degree of realism. As such, the Rahman-Stillinger papers of the early and mid 1970s had a huge impact. For the first time, scientists could see how water molecules organize and move. And literally, we did see, as Charles Bennett rendered a frequently viewed movie of liquid water from one of Anees Rahman's first trajectories. Soon after the early Rahman-Stillinger papers appeared, first ions (43), then hydrophobes (44) and then peptides (45) were added to the simulation boxes, setting the stage for modern computational modeling of aqueous solutions and bio-molecular simulations.

By computing correlation functions from their molecular simulations, Rahman and Stillinger established quantitative measures of motion and structure in liquid water. Their results showed that the correct picture is indeed that of a disordered percolating hydrogen-bonding network, locally tetrahedral, with motion dominated by switching hydrogen bond allegiances. In view of the simulation results, the two-state picture became utterly irrational, yet proponents have persisted with elaborate ideas on "virtual" (46) critical points, about which I have more to say later. (Observations of spectroscopic isosbestic points seemed to provide evidence to support some version of a two-state model, until Phillip Geissler (47) showed analytically that any inhomogeneously broadened line will exhibit isosbesticity, so the common interpretations of such phenomena are ill-conceived.)

## 8. Why treating water is a challenge for analytical theory

An underlying simplicity characterizes non-networked liquids like argon, benzene, acetonitrile and so forth. These are the liquids for which the WCA perspective is valid, where packing forces and entropy dominate intermolecular structure. At conditions where these liquids are homogeneous and without long-ranged order, density fluctuations are governed to an excellent approximation by a bi-linear free energy functional, implying that these fluctuations are the normal modes of the fluid. Density fluctuations in these materials therefore obey Gaussian statistics, which is amenable to analytical treatment (48). One consequence is the Percus-Yevick equation for the pair correlation function for hard sphere particles, which can be solved analytically (49). Another is its generalization for non-spherical particles called the "reference interaction site model" (i.e., the RISM equation) (50).

In contrast, water possesses local tetrahedral order, which propagates to macroscopic scales when the liquid freezes into ice. No such local remnant of crystal structure extending to second-nearest neighbors is recognizable in liquids where packing forces are dominant. In water, this local tetrahedral order is a consequence of directional hydrogen bonding, and it complicates steps that might be attempted when coarse-graining to produce a tractable theory. For example, upon integrating out orientational degrees of freedom, neighboring molecular centers will be coupled by no less than triples. Three (and more)-point functions are therefore intrinsic to water's interactions and its field-theoretic Hamiltonian. In this sense, Moore & Molinero's "mW" model (51) is a minimal model for simulating water as a liquid with local tetrahedral order. More to the point, a theoretically tractable Gaussian model, which is bi-linear, cannot describe significant molecular reorganization in water. Large fluctuations and non-linear response are therefore intrinsic to much of its microscopic behavior.

> Large fluctuations and non-linear response are intrinsic to much of water's microscopic behavior.

This issue has little to do with details of an intermolecular potential. For water, it is only important that the force field be one for which local tetrahedral ordering and a first-order liquid-ice transition are emergent. These features together with water's dipole moment are responsible for the most outstanding physical behaviors of water, and these features alone present a challenge to analytical treatments of the material. Work done to build more and more realistic force fields for classical simulations of liquid water, as useful as it might be, misses this point. Moreover, to the extent one wishes to improve quantitative accuracy of models of water, one must recognize that the molecule itself exhibits non-trivial intramolecular dynamics in its polarization and bond vibrations, and much of this dynamics along with librations are quantum mechanical.

Some interesting processes in water involve molecular reorganizations that are relatively small in amplitude, and a Gaussian or linear-response model can apply in those cases. Polarization fluctuations facilitating electron transfer is one example, where a Gaussian model yields Marcus theory of electron transfer (52). The model breaks down on small length scales where only a few water molecules contribute to polarization fluctuations — so-called "inner sphere" solvation. Starting in the late 1980s, liquid-state scientists put Marcus theory to a series of demanding tests. In particular, experiments and simulations studied solvation and dynamics of hydrated charge-transfer pairs, e.g., (53) and (54). Through measurements and calculations of distribution functions and response functions, the essential correctness of the Gaussian phenomenology was verified. I believe the results of these tests influenced the decision to award Rudy Marcus the 1992 Nobel Prize in Chemistry.

Density fluctuations accommodating hydration of small hydrophobic species is another example where a small-fluctuation Gaussian model applies. In this case it yields the Pratt-Chandler theory of hydrophobicity (55). (While the Gaussian nature of Marcus' theory was obvious immediately, the connection to a Gaussian field theory for Pratt-Chandler theory was not apparent until much later (48, 56).) This theory of hydrophocity breaks down at large length scales where hydrophobic interactions become strong and structure determining. The crossover to large fluctuations occurs because soft water-vapor interfaces nucleate when low-curvature hydrophobic surfaces extend beyond a nanometer (57).

It is then that surface tension of these interfaces controls forces of self assembly, as I came to understand for the first time with Ka Lum and John Weeks (58). Hydrophobicity on these nanometer scales, therefore, leverages the power of the liquid-vapor phase transition, even at temperatures well below the boiling temperature. It is a surface-induced pre-transition with assembling hydrophobic units manifesting, in effect, microscopic steam engines! (In recent work, my co-workers and I identified an analogous effect controlling membrane induced forces between proteins, which, in that case, were shown to leverag first-order transitions in lipid membranes (60).)

## 9. Rare events

Water's intrinsic chemistry adds another layer of complexity, such as reactions like $[H_2O]_{aq} \rightarrow [H]^+_{aq} + [OH]^-_{aq}$. I had not thought deeply about such process until one day in 1995, after completing a lecture in freshman chemistry on acid-base equilibrium. (It was my first experience being in charge of teaching Chem 1A, which I did with the help of my colleague Bill Miller. This teaching assignment raised concerns among some of our colleagues over whether two theorists were capable of teaching elementary chemistry. All indications at the end of the semester were that we did a fine job.) Returning to my office after that lecture, I pondered what could be the mechanism of this auto-ionization, the fundamental kinetic step of pH. What molecular reorganization could possibly facilitate or provoke a process by which a highly stable intact water molecule spontaneously becomes a hydroxide ion?

After reading papers by Manfred Eigen (61) and by my colleague Brad Moore (62), I came to appreciate that no one knew the answer to this question. Moreover, finding the answer would require a new development. In particular, I learned that the half-life of an intact water molecule in liquid water is about 10 hours, and when the auto-ionization takes place, it is completed in a few picoseconds or less. Without some special form of filtering or sampling, the probability of observing the event per molecule is therefore of order $10^{-16}$. Clearly, capturing the event in an experiment would be challenging.

Moreover, I thought, it would be no easier for theory. With the technology available in the 1990s (or today), no straightforward molecular simulation could elucidate the process because *ab initio* molecular dynamics was limited to trajectories of only a few picoseconds. Further, my earlier work on transition state theory and its dynamical correction (63) could not be applied because the relevant reaction coordinates and transition states were unknown. Indeed, no exploration of a potential energy landscape could locate transition states for the process because the number of saddles grows exponentially with system size, and transition states in a disordered system do not necessarily coincide with saddles. These difficulties are on top of the fact that *ab initio* methods can have difficulty estimating the relevant force field (64).

The theoretical puzzle presented by auto-ionization in water is quite general. It is the problem of simulating a rare but important event in cases where the mechanism cannot be known without observing trajectories. There are two common but incorrect ways to view such a process: assume it simply does not happen because it is not observed in a straightforward simulation of specific time duration; or assume the natural mechanism follows a pathway provoked by an artificial perturbation, like raising temperature or forcing a chosen coordinate. The former is obviously in error because the time scale for the rare event can be far longer than the time simulated; the latter is in error too, in this case because a complex system offers an exponentially huge variety of pathways, which makes it most unlikely that an arbitrarily chosen energy distribution or coordinate overlaps with physically relevant pathways.

> Rare event sampling has made possible molecular simulations of far-from-equilibrium phenomena, including chemical dynamics in water.

After thinking about this problem for a few years, I had the great luck in 1997 that Peter Bolhuis, Christoph Dellago and Phillip Geissler joined my research group. They helped me find a solution (65). The crucial insight leading to this development was the conception of sampling trajectory space in a fashion analogous to Monte Carlo sampling of state space. Specifically, we needed to recognize the possibility of taking random-walk steps in trajectory space (e.g., so-called "shooting" and "shifting" moves) with importance sampling achieved by accepting or rejecting those steps in accord with the prescribed distribution of trajectory space. In the case of auto-ionization in a deterministic model of water, that distribution is the equilibrium distribution of initial conditions subject to the constraints that *i.* a tagged water molecule is initially intact, and *ii.* that same water molecule gives up a proton to the surrounding liquid at the end of the trajectory.

We called this approach "transition path sampling" (TPS). It is an easy method to generalize and implement as a shell of commands around any molecular dynamics code, focusing attention on only those processes of interest. It is thus a means to edit trajectory space, to focus on important stretches of time, not unlike editing a movie, showing interesting parts, avoiding what is ordinary and boring. For water auto-ionization, the interesting parts are time frames surrounding the passage through a transition state, and a transition state is recognized from a statistical criterion — a configuration from which trajectories have equal probabilities of committing to the product or reactant. A narrated YouTube video (http://www.youtube.com/watch?v=zeFSzt5x9uo) shows one such trajectory of water autoionization harvested by TPS from an *ab initio* molecular dynamics developed by Michele Parrinello and his co-workers (66).

In addition to rare event applications, the TPS approach can be used to study nonequilibrium phase transitions through the computation of large-deviation functions (67), about which I have more to say shortly. In the decade since its initial formulation, many improvements and generalizations of TPS have appeared — transition interface sampling (68), forward flux sampling (69) and so forth — but the basic underlying idea remains the same: importance sampling of trajectory space (distinct from state space), without pre-conceived notions of mechanisms.

## 10. Fluctuation dominance

At the 2012 GRC on Water, David Limmer and I presented a preliminary report on using the TPS approach to treat water's glass transitions. This work, which was completed and published two years later (70), provided a theoretical rationalization of water's polyamorphism. "Polyamorphism" refers to multiple distinct amorphous phases. In water, such phases are glasses (i.e., amorphous ices). They are far-from-equilibrium states of matter, dominated by large fluctuations.

To appreciate how these states arise and why fluctuations are large and distinct from equilibrium behaviors, first recall that macroscopic samples of the liquid can be maintained at supercooled conditions for long periods of time, provided the samples are cooled carefully and not too far below the freezing temperature, 273 K. A small shake or seeding, of course, will cause ice to form quickly because those actions spur the formation of critical crystal nuclei. No matter how carefully it is supercooled, however, maintaining the liquid much below 235 K proves difficult. At those conditions, the probability of forming critical nuclei is large enough that ice emerges spontaneously from the liquid on a time scale of 1 s or less.

> Fluctuation dominance, where large deviations from the mean are controlling, occurs in water at deeply supercooled conditions.

Lowering the temperature yet further, to around 215 K, yields the fastest crystallization time scale, 0.1 ms. At that stage, the liquid is at its limit of metastability, and the material is possessed by fluctuations. Nanometer-scale domains of differing densities — ice-like and liquid-like densities — interconvert on time scales of tens of nanoseconds. Local regions with ice ordering can emerge in the lower-density domains, and where this ordering sets in, the low density is temporarily stabilized as a transient domain of crystal. Two or more such ordered regions can merge when they are similarly oriented but otherwise recede to a disordered and possibly higher density state. These random processes occur throughout, with tens of thousands of such processes completed before the thermodynamic driving force prevails, and macroscopic domains of crystal appear.

This dynamics — crystal coarsening — is beautiful to observe. It also seems easy to misinterpret. For example, a vivid YouTube presentation (https://www.youtube.com/watch?v=Y8jKM1b-jZU) shows molecular dynamics of simulated water forming ice after the stable liquid is suddenly quenched to its limit of stability. Throughout the non-equilibrium trajectory, fluctuations are large and ubiquitous. Those of the density, which are most pronounced in the early stages of the trajectory, have been interpreted as evidence of two-liquid coexistence, whereas the time scales and length scales of the fluctuations are those expected of coarsening (71). This idea of two distinct liquids and a metaphysical liquid-liquid critical point in supercooled water goes back more than two decades (72). Over those years, it has been discussed and debated at many GRCs and elsewhere. As Kurt Binder has pointed out (73), however, finite lifetimes of metastable states make it impossible for fluctuations of early-stage coarsening to be like criticality between two distinct liquid phases in pseudoequilibrium. The finite lifetime imposes a fundamental upper bound on possible correlation lengths so that standard critical phenomenology is inapplicable. Viewing ice coarsening on its own terms as a far-from-equilibrium behavior is straightforward by contrast (74).

Binder's observation illustrates the value of introspection. It reminds me of an observation made by Richard Feynman during an episode in the late 1960s and early 1970s. Some experiments were being interpreted to imply the existence of a new phase of liquid water called "polywater." It was supposed to be thermodynamically more stable than the normal liquid. No one any longer believes in polywater, but at that time, many did believe. Felix Franks has written about this period in water science (75), and he reports that Feynman said "There is no such thing as polywater because if there were, there would also be an animal which didn't need to eat food. It would just drink water and excrete polywater."

## 11. Glass transition

The rate of crystal ice formation no longer increases as temperature is lowered much below 215 K. That is because this cooling slows molecular motions and thus limits reorganization of disordered domains and misaligned neighboring micro crystal domains. Indeed, motions become so slow that cooling rapidly enough to a temperature far enough below 215 K will trap the material in a disordered solid state — a glass. The cooling rate to reach such an amorphous solid must overcome the fastest crystallization time, i.e., a cooling rate of at least $10^2 \, \text{K}/0.1 \, \text{ms} = 10^6 \, \text{K/s}$ (76). The process of making this glass is therefore path dependent (i.e, irreversible) and far from equilibrium.

Given how molecular reorganization is arrested by cooling, it is clear why the material is fluctuation dominated. Many different microscopic domains are trapped in configurations that differ from an average equilibrium configuration. Moreover, it is clear why cooling at yet faster rates will produce less stable glasses. The ultimate fate of glass is to become crystal. The more stable the glass the slower the conversion from glass to crystal, and reaching such a state of metastability requires time for local domains to anneal. Simply dropping the temperature instantly from that of a liquid will produce an especially poor glass because domains will have no chance to anneal, and it will quickly age and convert to crystal.

Competing time scales are common to all glass forming materials. Water exhibits this competition, but it is not a particularly good glass former because its fastest crystallization times are relatively short. Nevertheless, it has the interesting (though not unique) additional feature that it can exhibit seemingly distinct glassy states. These states, accessed by different paths manipulating rates of change in both pressure and temperature (77), are the low-density and high-density amorphous ices. The former (LDA) has the higher fraction of crystal-like ordered domains, while the latter (HDA) has the higher fraction of disordered domains. Some have viewed this polyamorphism as a smooth continuation of distinct (though unobservable) liquid phases. The fundamental problem with such a picture is that a singularity or broken symmetry exists, which partitions ergodic liquid behavior from non-ergodic solid behavior. A smooth continuation does not exist in such situations. When the solid is a glass, this broken symmetry is manifested out of equilibrium, in trajectory space. Coexistence between two amorphous ices can exist, but it necessarily ends with a singularity where the solids melt — a non-equilibrium analog of a triple point, not a critical point.

> Glass is not simply an extrapolated liquid because the glass transition manifests a broken symmetry and an associated singularity in the partition function sum over trajectories.

Broken symmetry in trajectory space is analyzed in terms of partition functions of trajectories — sums over histories (78). In equilibrium statistical mechanics, phase transitions are analyzed in terms of free energy, and free energy is the logarithm of the partition function sum over micro states. In non-equilibrium theory, the analog of free energy is the large deviation function. There are various algorithms that can be used to efficiently sum over histories and compute large deviation functions, and TPS is one such method. TPS was initially conceived for cases where large deviations from the mean are rare events that are localized in time. Yet it can also be applied in certain cases as an importance sampling of long trajectories. In the case of glass transitions, they are trajectories that appear to be non-ergodic, which are highly improbable in the equilibrium ensemble of trajectories. They are trajectories where configurations continue to overlap with initial conditions over long periods of time.

Mauro Merrolle, Juan Garrahan and I used TPS in the first demonstration that glass transitions can be analyzed in this way in terms of large deviation theory (79). It was therefore the natural approach for David Limmer and me to take in our investigation of amorphous ices (70). Carrying out TPS simulations for one model of water, we were able to determine glass-transition temperatures as functions of pressure, examine path dependence of different amorphous states, provide analytical theory for boundaries separating these states, and interpret several experiments. While numerous improvements are possible — improved models, larger system sizes, longer trajectories, and so forth — the methodology shows the way for interpreting this class of far-from-equilibrium behavior.

## 12. Epilogue

There is a direct line between these recent developments on glass transitions and the work I was first exposed to during and around the time of my first GRC in 1967 — the modern theory of phase transitions, non-equilibrium processes, and molecular simulation. There is also a style to this work that I trace to that time — striving for rigorous analysis and physical principles that underlie models applied to observations. It is a perspective that often leads me instinctively to view natural phenomena in terms of cooperativity, interfaces and large fluctuations. It also leads me to dismiss explanations constructed from hypothetical but unobservable phenomena and to distrust explanations built from extrapolation.

I recall Michael Fisher in the early 1970s complaining about the work of a theorist who had extrapolated the Hartree theory to interpret a phase transition. Hartree theory is an example of a self-consistent Gaussian fluctuation approximation; it can be valid when fluctuations are small, but it is divergent when extrapolated too far outside that regime. Fisher remarked, "Isn't it overly optimistic to attribute the breakdown of a theory to the breakdown of a physical system?" The renormalization group theory and surrounding developments ended the popularity of such extrapolations in equilibrium statistical mechanics. Conditions out of equilibrium can be a different story. In current times, for example, Fisher's comment could be leveled at applications of mode coupling theory (another small fluctuation approximation) to the glass transition. And there is still no consensus on the glass transition.

Valuable information about glass transitions has been gleaned both from exact solutions of idealized models and from numerical work like calculations of large deviation functions, but mystery persists. I expect the topic will be a focus of interesting research for a few more years. One challenge is the competition with coarsening, as discussed above. Another is more fundamental — the cooperative nature of dynamics, where domains of moving atoms are localized and segregated from those that are jammed. It is understood how this dynamical heterogeneity is essential to hierarchical energy scales and length scales characterizing relaxation of glass formers, but a compelling and general molecular theory for how this heterogeneity emerges remains unknown.

> Far from equilibrium, large fluctuations and interfaces are features of the interesting topics of contemporary and future liquid state science.

Being far from equilibrium, exhibiting large fluctuations and inhomogeneity, glassy physics seems prototypical of most topics meriting attention in contemporary liquid state science. A scan of a recent Liquids GRC program (e.g., https://www.grc.org/programs.aspx?id=11489 ) confirms this impression. This broad field of research is now sometimes called the study of "soft matter" and sometimes called the study of "complex fluids." It links physics and chemistry to biology and materials science. It is the general study of condensed disordered systems, which are sometimes fluid, sometimes gel, sometimes glass, sometimes granular, even sometimes alive. In each of the sub areas, I anticipate significant progress, all built upon the scientific foundation that was first being assembled 50 years ago. I feel rewarded to have witnessed the building of that foundation and to have participated in the advances during these last 50 years.

## DISCLOSURE STATEMENT

The author is not aware of any affiliations, memberships, funding, or financial holdings that might be perceived as affecting the objectivity of this review.

## ACKNOWLEDGMENTS

I have greatly benefited from working with students and colleagues at the University of Illinois, Urbana-Champaign (1970–83), at the University of Pennsylvania (1983–86), and at the University of California, Berkeley (1986–present). Throughout, I have been fortunate to have nearly continuous support from the National Science Foundation, occasional support from the National Institutes of Health, and nearly three decades of support from the Basic Energy Sciences Division of the Department of Energy. In that latter case, some of that funding came to me through association with Lawrence Berkeley National Laboratory. I am grateful for all these associations and support.